\documentstyle[twocolumn,pre,aps,epsf]{revtex}

\newcommand{\be}{\begin{equation}}
\newcommand{\ee}{\end{equation}}

\begin{document}
\draft

\twocolumn[\hsize\textwidth\columnwidth\hsize\csname @twocolumnfalse\endcsname

\title{Phase transition between $d$-wave and  anisotropic $s$-wave gaps
in   high temperature oxides superconductors }
\author{Iksoo Chang${}^{1}$, Jacques Friedel${}^2$, and
Mahito Kohmoto${}^3$ }
\address{${}^1$Department of Physics, Pusan National University, Pusan
609-735, Korea}
\address{${}^2$Laboratoir\`{e} de Physique des Solides, Universit\'{e}
Paris-Sud,
 Centre d'Orsay, 91405 Orsay Cedex, France}
\address{${}$ unit associated to the CNRS}
\address{${}^3$Institute for Solid State Physics,
University of Tokyo, 7-22-1 Roppongi, Minato-ku, Tokyo, Japan}

\maketitle
\begin{abstract}
We study models for superconductivity with two  interactions:  $V^>$ due to
antiferromagnetic(AF) fluctuations
 and $V^<$ due to phonons, in a weak coupling approach to the
high temperature superconductivity. The nature of the two interactions are
considerably different
; $V^>$ is positive and  sharply peaked at ($\pm\pi$,$ \pm\pi$)
while
$V^<$ is negative and  peaked at ($0,0$) due to weak phonon
screening. We numerically find
(a) weak BCS attraction is enough to have high critical temperature if a van
Hove
anomaly is at work,
(b) $V^>$ (AF) is important to give $d$-wave superconductivity,
(c) the gap order parameter $\Delta({\bf k})$ is constant($s$-wave) at
extremely
overdope region and it  changes to anisotropic $s$-wave as doping is
reduced,
(d) there exists a  first order phase transition between  $d$-wave and
anisotropic
$s$-wave gaps.
These results are qualitatively in agreement with preceding works; they
should be
modified in the strongly underdope region by the presence of antiferromagnetic
fluctuations and ensuing AF pseudogap.
\end{abstract}
\pacs{ 74.20.-z, 74.20.Fg}
]

\narrowtext

One of the most significant works in recent condensed matter physics is
undoubtedly
the  discovery of high $T_c$ copper oxides\cite{BM}.
It seems the high value of $T_c$ can not be
understood by the 'classical' BCS theory\cite{BCS}, a large amount of
experimental and theoretical works have been performed. In fact, quite a few
number of high $T_c$ superconductors(HTSC) are found.

A phase diagram with doping and temperature is
very rich, with the
antiferromagnetic ordered phase, the superconductive phases -- underdope and
overdope--, a pseudogap region as well as a 'normal' metallic range.
 Electron doped
compounds show a similar succession of AF, superconductive and normal metallic
phases, with lower critical temperatures. It is notable that,
in common with the classical BCS case, the superconductive state occurs
through a condensation of cooper singlet pairing $({\bf k}\uparrow, -{\bf
k}\downarrow)$.

Currently  there are many works that ignore  phonon-mediated
interactions and, in many cases,  van Hove anomaly altogether. In these
studies one considers mostly interaction due to AF fluctuations. The main
justification for these approaches is that the energy scale for magnetic
interaction
is $\simeq 1000K$ which is somewhat larger than the Debye temperature of
$\simeq
300K$. In fact, the solutions of the gap equation with only positive
interaction
due to AF fluctuations gives $T_c \simeq 100K$ \cite{moriya,pines} which is
consistent with the  experimental results. However it is not surprising
since $T_c$
varies a lot depending on the coupling constants.

On the other hand there are
a number of works
\cite{hs,friedel,newns,abrikosov,bok} which show enhancement of
$T_c$ due to a van Hove anomaly. In most of these investigations, the main
purpose
is to obtain high critical temperature which is consistent with the $T_c$
observed
at optimal doping. Therefore almost all the works except ref.[6]
assume that
a van Hove anomaly is on the Fermi surface at optimal doping. In this way the
decrease of
$T_c$ from the optimal as doping is increased or decreased is natural.
However, the
systematic computation of doping dependence has not been done. A potentially
serious
problem of these approaches is that antiferromagnetic (AF) fluctuations  which
should be important in the underdoped region are not taken into account.

Recently, Friedel and Kohmoto\cite{fk} consider the case where
the van Hove anomaly  is at 'Fermi level'
for undoped(half-filling). Besides phonon mediated interaction, one from AF
fluctuations is considered.
Indeed, it was shown the AF instabilities decrease faster
with increasing doping than the effect of the van Hove anomaly on
superconductivity\cite{friedel}. They
argue the followings: Phonon or electron
mediated weak BCS attraction is enough to have high critical temperature if
a van
Hove anomaly is at work near enough to the Fermi level. This could apply to
electron
doped compounds and also to compounds with CuO$_2$ planes overdope in holes,
where
$T_c$ decreases with increasing doping. If phonons dominate, it should lead
to an
anisotropic but mainly
$s$ superconductive gap and probably
also in electron doped compounds. A
$d$ gap should develop as observed in a number of cases if AF fluctuations are
strong enough. In the underdope range, the observed decrease of $T_c$ with hole
doping can be related in all cases to the development of antiferromagnetic
fluctuations which produces a magnetic pseudogap, thus lowering the density of
states at the Fermi level. The observed mainly $d$ superconductive gap then
can be
attributed to antiferromagnetic fluctuations.

The essential feature of the BCS theory is the pairing picture. This has
been well
established in the 'classical' superconductors. Also there is almost no sign of
breakdown of this picture  in HTSC. However, in
order to obtain the so called 'BCS formula', one needs  the approximations:~~
(a) The pairing interaction is weak,
(b) The density of states is not too fast varying near the Fermi surface,
(c) The pairing interaction is constant within
 the cutoffs
$\pm k_B T_D$ near the Fermi surface and zero otherwise.
These approximations lead to the BCS formula:
 $T_c \simeq 1.13 k_B T_D \exp\{-1 / N(0)V\}$,
$\Delta(T=0)/ k_BT_c \simeq 1.75$, the universal  specific heat jump  at $T_c$,
the
isotope effect with exponent
$\alpha = 1/ 2$ etc.

We assume (a) throughout  this work, but (b) is certainly not appropriate as
shown
below. There is a large enhancement of the density of states close to the
van Hove
anomalies. The approximation (c) is oversimplified  and  will be
replaced by weakly screened phonon mediated interaction, by AF repulsive
interaction, and in addition by both. Therefore we expect that
the BCS formula can be violated in experiments, even if the BCS theory with
a singlet pairing is correct.

In order to examine the  above ideas, we consider a model on the square
lattice with
specific interactions between electrons and solve the gap equation for general
values of hole doping. The
effects of
pseudogap are {\em not} taken into account nor the correction due to the small
couplings between CuO$_2$ planes and the ensuing fluctuations of
superconductivity
near $T_c$. Then
$T_c$ keeps rising while doping decreases. This contrasts with 
the behavior of HTSC
whose $T_c$ decreases and eventually vanishes in underdope region.

The high $T_c$ oxides have a square lattice of Cu atoms and O atoms in between
the Cu
atoms (here we neglect the orthorombic distortion which exists for some
HTSC). We
take an effective tight-binding picture to consider the Cu sites only, then
with
effective transfer $t$, we have the following results.

For undoped compounds, the Fermi level is then a square. Doping by
electrons and by holes produce almost square Fermi surfaces with nearly
symmetric
deviations from the undoped square surface. This fundamental symmetry between
electron and hole dopings fits well the general symmetry of the phase diagram
observed for electron and for hole doped compounds.

As a result of this geometry the Fermi
level sits also near a strong peak in the density of states, which diverges
logarithmic at ($\pm \pi, 0$), ($0,\pm \pi$) for the undoped  compounds.
This van Hove anomaly is characteristic of the (quasi) two-dimensional
compounds, where it is much stronger than in more isotropic $3d$ compounds.

The kinetic part of the square lattice model with only nearest-neighbor
hopping can
be written
$\varepsilon (k) \simeq -2t(\cos k_x  + \cos k_y )$
where the transfer $t$ is typically $0.25$eV and energy will be
measured in
units of $t$ from now on. In this energy dispersion, there are two saddle
points
which lead to the van Hove singularties at ($\pm \pi,0$)
and ($0,\pm \pi$).
The gap equation is

\be
\Delta({\bf k}) =
-{1 \over 2}\sum_{k'} V_{{\bf {kk'}}} \Delta({\bf k'}) \tanh({E_{k'} \over 2k_B
T})\, {1\over E_{k'}},  \label{gap}
\ee
where  $\Delta({\bf k})$ is the gap order
parameter,
$V_{{\bf {kk'}}}$ is the interaction between electrons, and
$E_k =\sqrt{\{\varepsilon (k)-\mu\}^2 + \Delta({\bf k})^2} $
where $\mu$ is the chemical potential. \\

--{\it BCS isotropic interaction $V_{{\bf {kk'}}} = {\rm const.}
(=-3t) < 0$ }.
In the overdope range
one can reasonably neglect the  effects of AF fluctuations in the
'normal'
metallic range. A high critical temperature
$T_c$ can then be obtained in the mean field BCS approximation.

The critical temperature is plotted in Fig.~1 for
weak coupling region where $N |V| \leq 0.3$ where $N$ is DOS near the bottom
of the
band. At the bottom of the band $T_c \simeq 10K$ which is quite reasonable
for weak
coupling BCS theory. It has an
 exact particle-hole
duality. In the extremely overdope region, the BCS prediction is well
satisfied because DOS is not so fast varying within the
cutoffs. This is no longer true for lower dopings where DOS is larger
and changes rapidly.
The enhancement of
$T_c$ is considerable there, about 150K at undoped.   The gap is
s-wave since the interaction is constant. This result shows that weak BCS
interaction is enough to have high $T_c$.

 We stress that this enhancement of
$T_c$ is only induced by  the DOS effect. The coupling does not play any
role. It seems that it is not totally unrealistic to expect extremely high
$T_c$ like  room temperature. However, there exist  AF effects near
undoped in the all known high
$T_c$ superconductors. It is likely to suppresses $T_c$. \\

\begin{figure}
\narrowtext
\centerline{\hbox{\epsfysize=2.5in \epsffile{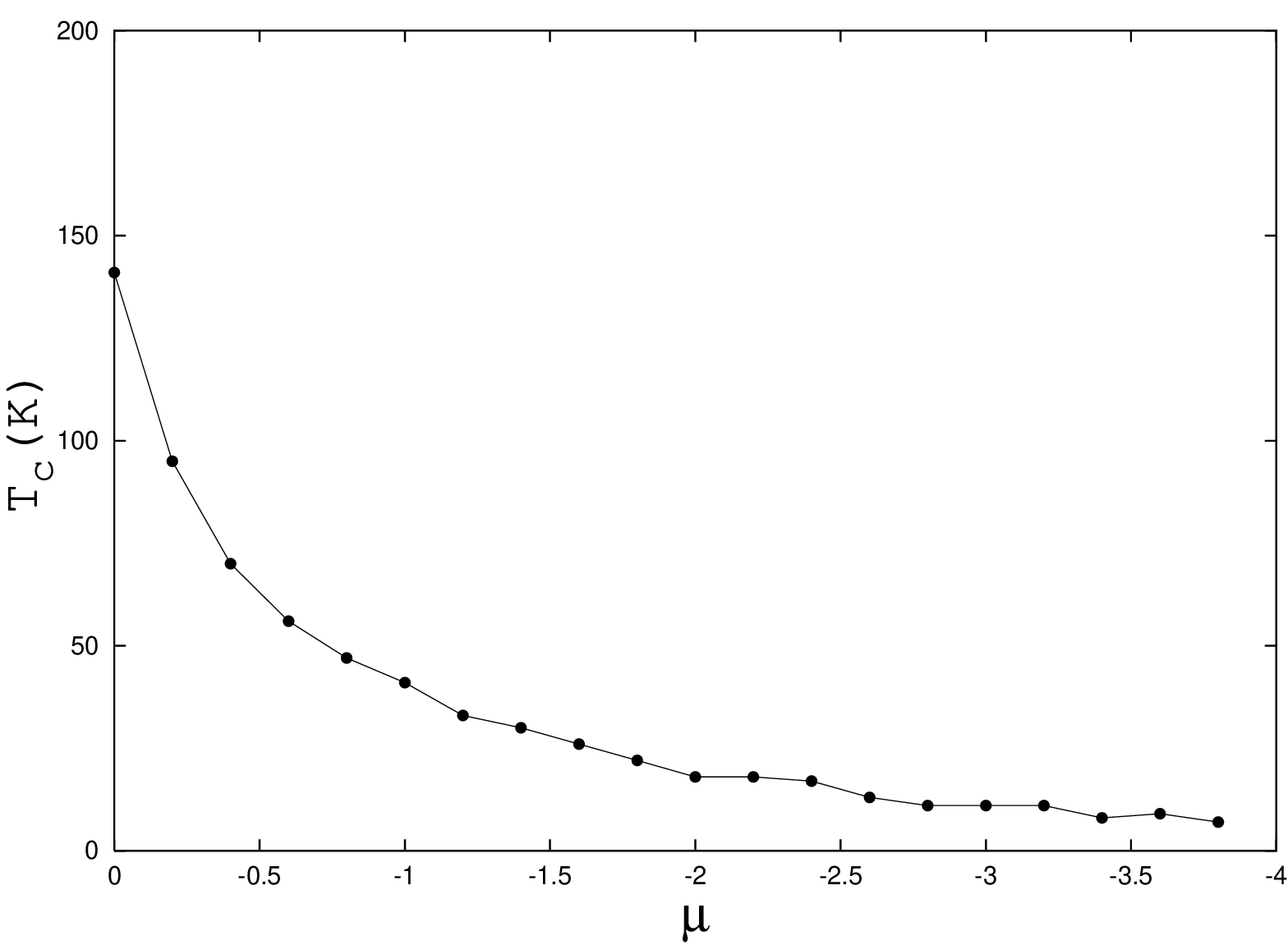}}}
\caption{\protect\footnotesize
Critical temperature $T_c$ vs chemical potential for BCS attractive
interaction ($V_{{\bf {kk'}}} = -3t <0$).
}
\end{figure}

-- {\it Weakly screened phonon-mediated interaction $V^<$}.  Let us
consider the
interaction
\be
V^<_{{\bf k},{\bf k'}} = -{|g_{\bf q}|^2 \over {|{\bf q}|^2 + |{\bf q}_0|^2}},
\label{v1}
\ee
where ${\bf q} = {\bf k}-{\bf k'}$,
$ g_q$ is the electron-phonon coupling and $2 \pi /{\bf q}_0$ is the
screening length. This type of phonon contribution is used in
ref.[8,9].
We take a large screening radius about 30 lattice spacing because of poor 2d
screening. The sign of
$V^<_{{\bf k},{\bf k}'}$ is always negative.

Critical temperature is displayed in Fig.~2.
The gap $\Delta({\bf k})$ now has   an extended $s$-wave symmetry near
half-filling. The new feature is that the gap not only depends on
the angle of
${\bf k}$ but also on the absolute value $|{\bf k}|$ considerably being
maximum at
Fermi level. See Fig.2(a),(b), and (c).
Possibility of this phenomena has not been considered seriously
in the
past. In most of the previous  works on superconductivity,
$\Delta({\bf k})$ is computed for ${\bf k}$ at the Fermi level.

A notable feature is that $T_c$ does not decrease by doping very much. In this
region it is very hard to have pure HTSC samples. Thus it is likely that our
results do not apply.
For extremely doped region
we have also
high $T_c$ superconductivity. This behavior will be discussed
elsewhere\cite{wo}.\\

\begin{figure}
\narrowtext
\centerline{\hbox{\epsfysize=2.5in \epsffile{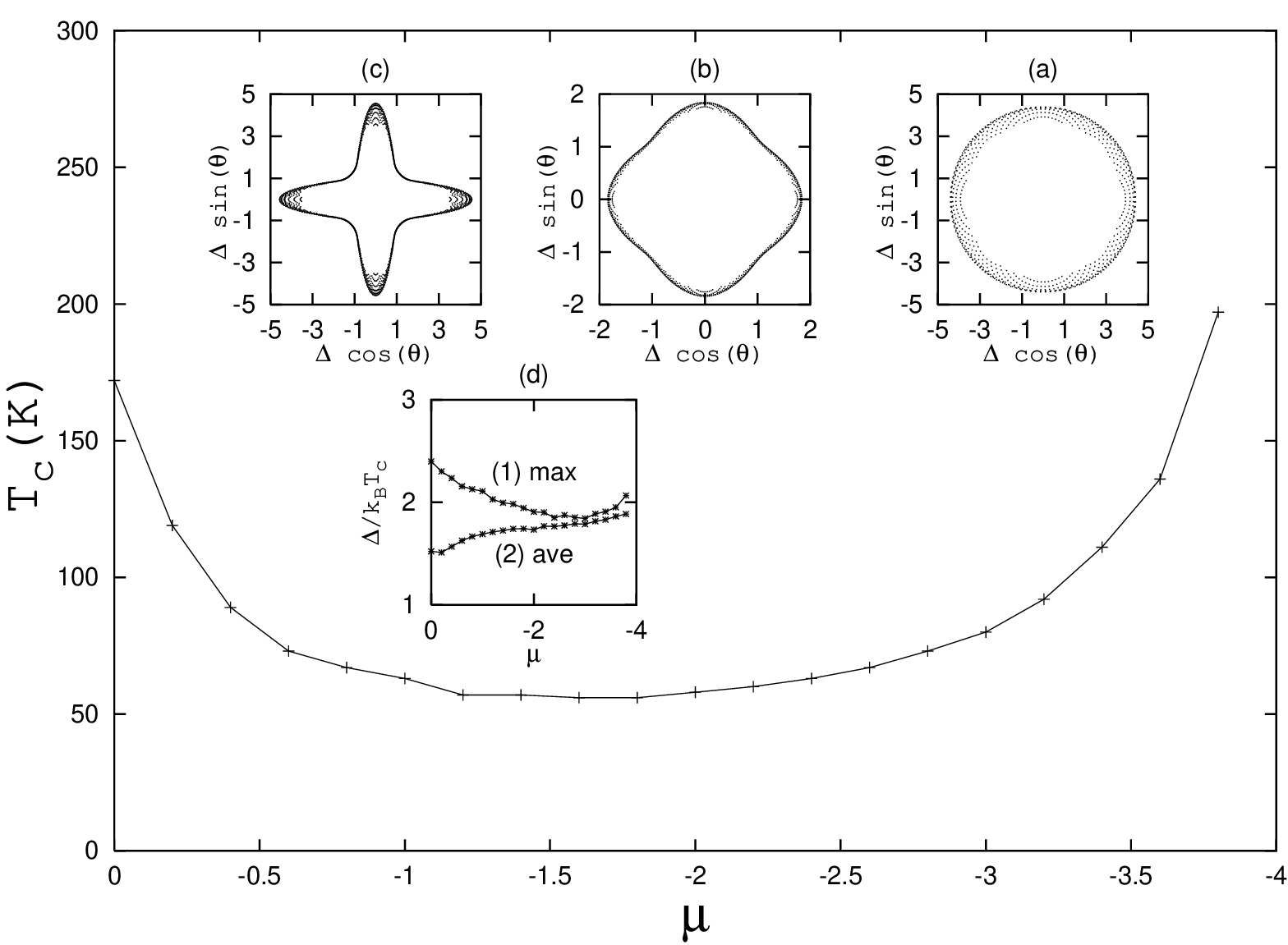}}}
\caption{\protect\footnotesize
Critical temperature $T_c$ for weakly screened phonon-mediated interaction
($V^<_{{\bf k},{\bf k'}} = -|g_q|^2 / ({\bf q}^2 + {\bf q_0}^2)$), and
polar plot of gaps at
$T=0$ for
(a) $\mu = -3.6$ (extended isotropic s-gap),
(b) $\mu = -2.0$ (slightly anisotropic s-gap), and
(c) $\mu = -0.2$ (anisotropic s-gap).
The distance from origin is the amplitude of gap for a given angle
$\theta = tan^{-1}(k_y/k_x)$.
(d) (1)$\Delta (0)_{max} / k_BT_c$ and (2)$\Delta (0)_{ave} / k_BT_c$ vs
$\mu$.
}
\end{figure}

 --{\it AF fluctuation mediated interaction
$V^>$}.   There are a number of attempt to
obtain high $T_c$ within the weak coupling BCS theory only from repulsive
interactions which are originated from AF fluctuations \cite{moriya,pines}.
Here we take the form of interaction by Monthoux et al. \cite{pines}
This interaction has peaks at $(\pm \pi, \pm \pi)$. We, however, take much
weaker
coupling about $1/10$ compared with one used by Monthoux et al. \cite{pines}
(The
peak value of their interaction is about $80eV$). We obtain  relatively low
$T_c$ ($T_c \sim 30 K$ at undoped)
and it vanishes at doping $c \simeq 0.1$.
The result obtained by using the
coupling of
ref.[4] is also confirmed.

This contrasts with attractive interaction which
always give superconductivity at low temperature for all doping.
The gap is always
$d$-wave for this type of repulsive interaction \cite{fk}.
Also we do not obtain a gap which breaks the square symmetry like the one
studied by
ref.[12] even in low temperatures. \\

--{\it The two types of interaction $V^> + V^<$ }.  Finally we take both
interactions into account. The two interactions do not cancel directly since
$V^>$
has peaks at ($\pm \pi,\pm \pi$) and $V^<$  at ($0,0$). In any event,  the
peak of
$V^>_{{\bf k},{\bf k}'}$ is fixed at
$(\pm\pi,\pm\pi )$ and the effect of the interaction near Fermi level strongly
depends on doping.  On the other hand
$V^<_{{\bf k},{\bf k}'}$ does not so much depend on doping. Thus the relative
strength of the two interactions changes with doping.

The  results are shown in Fig.3.
In case both s and d gap can comprise the (mixed) solution of gap equation
for a given point $(\mu, T)$ in phase space, the  gap function
is chosen as the one which has the lower free energy.
A notable feature is that, although
$V^>$ is not more effective for superconductivity, it is very effective for
$d$ wave
gap. We find a first order transition line between $d$ and
anisotropic
$s$-wave gap. The latent heat is also displayed in Fig.3(d).
There are experimental results which show a nodeless
HTSC\cite{deutscher1,deutscher2}.

\begin{figure}
\narrowtext
\centerline{\hbox{\epsfysize=2.5in \epsffile{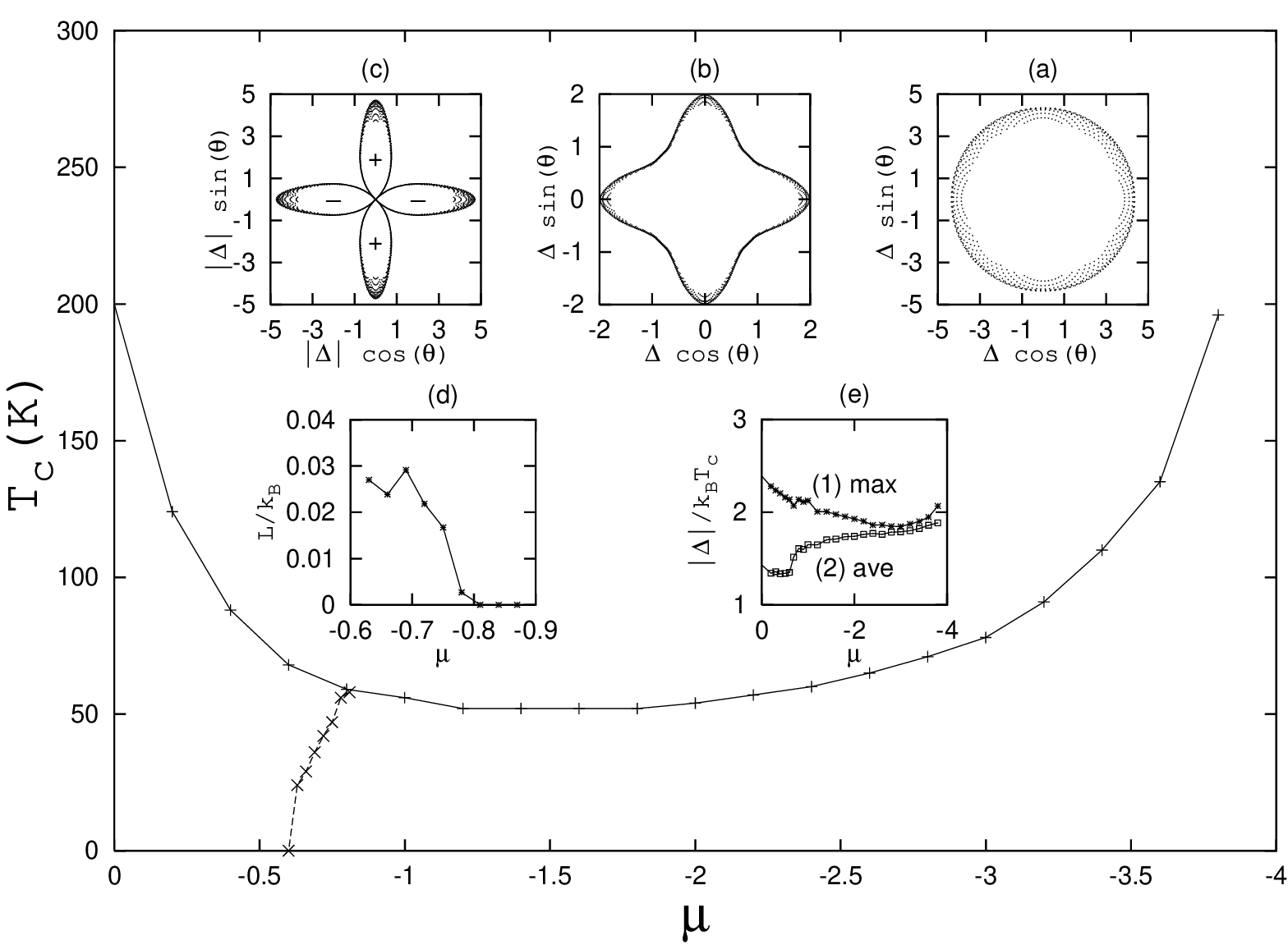}}}
\caption{\protect\footnotesize
Critical temperatures $T_c$ and the first order phase transition line
(dotted line) for the interaction
$ V^< + V^>$, and polar plot of gaps at
$T=0$ for
(a) $\mu = -3.6$ (extended isotropic s-gap),
(b) $\mu = -1.0$ (anisotropic s-gap), and
(c) $\mu = -0.2$ (d-gap).
(d) is the latent heat across the first order transition line.
(e) (1)$|\Delta (0)|_{max} / k_BT_c$ and
    (2)$|\Delta (0)|_{ave} / k_BT_c$ vs $\mu$.
}
\end{figure}

The quantity $|\Delta (0)|_{max} / k_BT_c$ takes the usual BCS value about
1.75 for
isotropic region (See Fig.3(e1)).
In the anisotropic region it is larger and about 2.5 which is not
surprising because the gap is anisotropic and even has nodes in $d$-wave
cases. This ratio with $|\Delta (0)|_{ave}$ averaged over ${\bf k}$ is
also plotted
in Fig.3(e2).

Our model has the particle-hole symmetry, thus we expect that the
electron-doped
compounds have similar gap structure. Mostly $s$-gap reported may be
anisotropic
and there may be a region of $d$-wave pairing in underdope region\cite{maeda}.

We also find a novel scaling relation; Namely, the absolute value of
average gap function $|\Delta (T)|$
as a function of $T$ for different chemical potentials can be
collapsed into  a universal curve by

\be
1 - {|\Delta (T)| \over |\Delta (0)|} = f(t/t_{char})
\ee
where $t = 1 - T / T_c$ is the reduced temperature and
$t_{char}$ is the characteristic reduced temperature for the decay of
$1 - |\Delta (T)| / |\Delta (0)|$ as $t$ increases for different chemical
potentials, and
$f$ is a universal scaling function.
Figure 4 shows a good scaling collapse of the rescaled gap functions
in terms of $t/t_{char}$ for different chemical potentials.
We observe that $|\Delta (T)|$ decreases as $T$ increases by one fashion
for $\mu$ below $-1.4$ and by the other fashion for $\mu$ above $-1.4$
(See fig.4(a)).
In fact Fig.4(b) shows that there are two different scaling
functions for $\mu$ below and above $-1.4$ at which $T_c$ becomes minimum.
This universal scaling does not apply for the region of first order
phase transition.

\begin{figure}
\narrowtext
\centerline{\hbox{\epsfysize=2.5in \epsffile{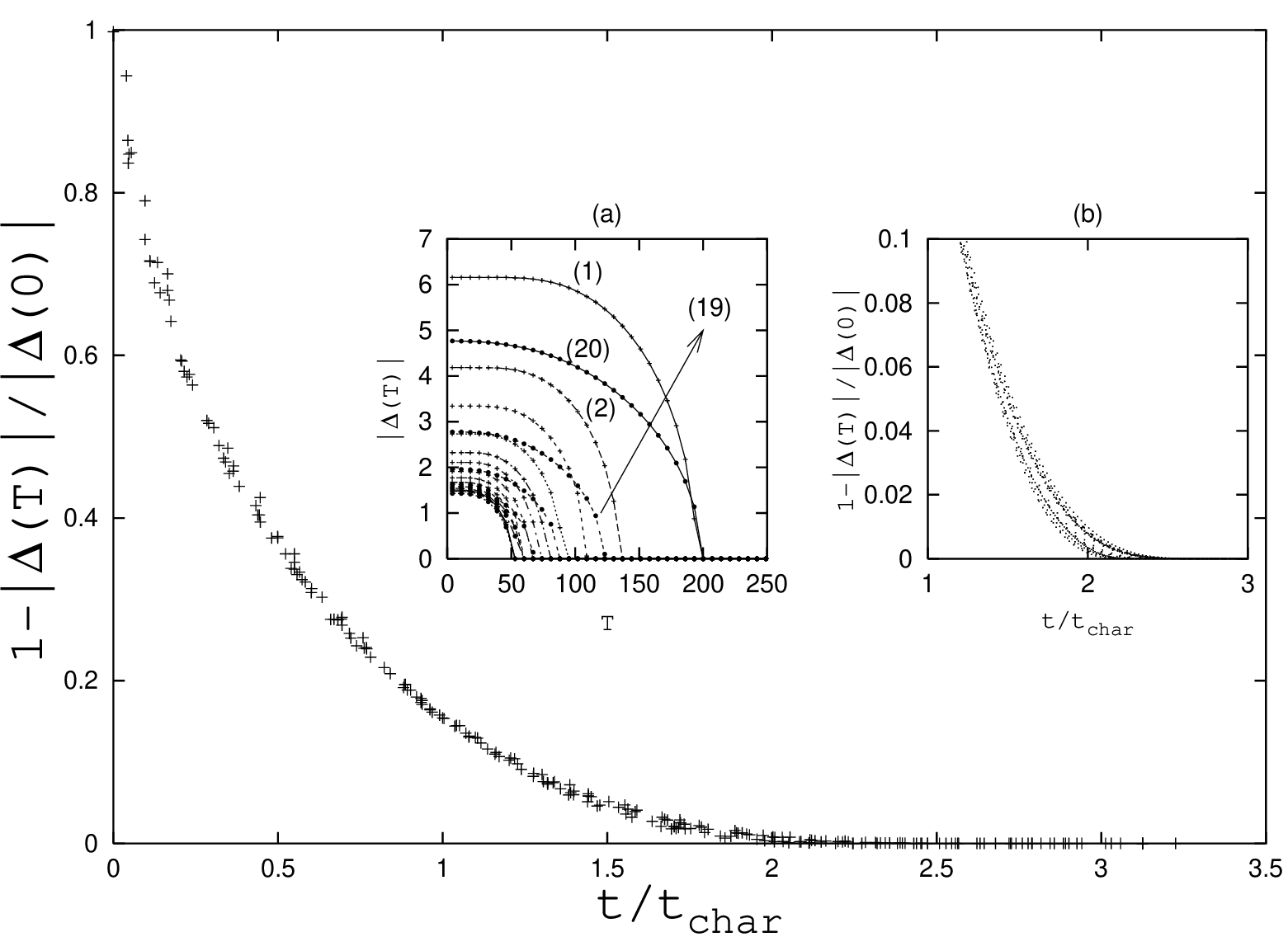}}}
\caption{\protect\footnotesize
The scaling collapse of rescaled gap functions versus $t / t_{char}$
for different chemical potentials from $\mu = -4.0$ to $\mu = 0.0$.
The inset $(a)$ shows $|\Delta (T)|$ vs $T$ for different chemical potentials,
which show different decaying behavior as $T$ below
and above $\mu = -1.4$.
The curves $(1), (2)$ for $\mu=-3.8, -3.6$ decay in the same fashion, but
differently from that of the curves $(19), (20)$ for $\mu=-0.2, 0.0$.
The inset $(b)$
shows that there are two different scaling functions below
and above $\mu = -1.4$.
}
\end{figure}

To conclude,
the presence of totally different types of interactions can be expected to
lead to
rich physics. 

We find that the phonon-mediated interaction is enough to have high
temperature superconductivity.  The role of AF fluctuation is to have
the d-wave gap. More importantly it gives the AF pseudogap which we do
not   take into account by technical difficulties. Probably the most
important prediction is the existence of the first order phase transition
betwen d and anisotropic s gaps which may be tested in experiments with
clean samples in the overdope region.

\begin {center}
{\large{\bf Acknowledgments}}
\end {center}

It is a pleasure to thank M.T. B\'{e}al, B.I. Halperin, Y. Takada,
J. Shiraishi for
useful discussions. M.K is grateful for the hospitability at Institut Henri
Poincare, Paris and Department of Physics, Pusan National University where a
part of this paper was written. I.C is grateful to Korea-Japan binational
program through KOSEF-JSPS which made his visit to ISSP, Tokyo possible,
to SRC program of KOSEF through RCDAMP, and to BSRI98-2412 of Ministry of
Education, Korea.


\begin{references}

\bibitem{BM}  L.G. Bednorz and K.A. M\"{u}ller, Z. Phys. B{\bf 64}, 199 (1986).

\bibitem{BCS} J. Bardeen, L.N. Cooper, J.R. Schrieffer, Phys. Rev. {\bf
106}, 162;
{\bf 108}, 1175 (1957).

\bibitem{moriya} T. Moriya, Y. Takahashi, and K. Ueda, J. Phys. Soc. Jpn. {\bf
59}, 2905 (1990); T. Moriya and K. Ueda, J. Phys. Soc. Jpn. {\bf 63}, 1871
(1994).

\bibitem{pines} P. Monthoux, A.V. Balatsky, and D. Pines, Phys. Rev. Lett. {\bf
67}, 3448 (1991); P. Monthoux, A.V. Balatsky, and D. Pines, Phys. Rev. B {\bf
46}, 114803 (1992).

\bibitem{hs} J.E. Hirsch and
D.J. Scalapino, Phys. Rev. Lett. {\bf 56}, 2732 (1986).

\bibitem{friedel} J. Friedel, Physica C, {\bf 153-5}, 1610(1988); J.
Phys: Condens. Matter. {\bf 1}, 7757 (1989); Nato Institute on Condensed
Matter,
Biarritz (1990).

\bibitem{newns} D.M. Newns, C.C. Tuei, P.C. Pattnaik and C.L. Kane, Comments
Cond.
Mat. Physics {\bf 15} 273 (1992).

\bibitem{abrikosov} A.A. Abrikosov, Physica C{\bf 222}, 191 (1994); {\it ibid}.
C{\bf 244}, 243 (1996).

\bibitem{bok} J. Bouvier and J. Bok, Physica C {\bf 249}, 117 (1995); '{\it Gap
Symmetry and Fluctuations in High $T_c$ Superconductors}' ed. J. Bok, G.
Deutscher
, D. Pavunaand and S.A. Wolf, Plenum Press, New York (1998)  NATO ASI Series,
Series B Physics {\bf 371} 77 .

\bibitem{fk} J. Friedel and M. Kohmoto, (preprint).

\bibitem{wo} Mahito Kohmoto and Iksoo Chang, (unpublished).

\bibitem{beal} M.T. B\'{e}al and K.Maki, Phys. Rev. B{\bf 53}, 5775 (1996); R.
Nemetschek et al., The European Phys. J. {\bf 5}, 495 (1998).







\bibitem{deutscher1} G. Deutscher, N. Achsaf, D. Goldschmidt, A
Revcholevski, A.
Vietkin, Physica C{\bf 282-287}, 140 (1997).

\bibitem{deutscher2} G. Deutscher and R. Maynard, {\it The Gap Symmetly
and Fluctuation in High
$T_C$ Superconductors} edited by J. Bok, G. Deutscher, D. Pavnua  and S.A. Wolf
NATO ASI Series, Series B Physics {\bf 371} (Plenum Press, New York 1998)
p.15; {\it ibid}. p.503,
NATO Advanced Study Institute Carg\`{e}ses (1997), ed. J.Bok and Pavuna, Plenum
Press London(1998) under press.

\bibitem{maeda} A. Maeda, H. Yasuda and T. Hanaguri, J. Phys. Soc. Jpn.{\bf
68},
594 (1999).







\end{references}
\end{document}